\documentclass[11pt]{article}
\usepackage[english]{babel}
\input{epsf}

\def\be{\begin{equation}}
\def\ee{\end{equation}}
\def\ba{\begin{eqnarray}}
\def\ea{\end{eqnarray}}

\def\12{{1\over 2}}

\def\msun{M_\odot}

\def\ltsima{$\; \buildrel < \over \sim \;$}
\def\simlt{\lower.5ex\hbox{\ltsima}}
\def\gtsima{$\; \buildrel > \over \sim \;$}
\def\simgt{\lower.5ex\hbox{\gtsima}}

\def\tento#1{\times 10^{#1}}
\def\Tg{T_\gamma}

\begin{document}
\selectlanguage{english}

\title{\bf The Influence of Ultra-High-Energy Cosmic Rays on Star 
Formation in the Early Universe}
\author{E.~O.~Vasiliev and Yu.~A.~Shchekinov\thanks{yus@phys.rsu.ru}\\
\it Rostov State University, Rostov-on-Don, Russia \\
{\small Received July 3, 2004; in final form, April 14, 2006} }

\date{\small Astronomy Reports, 2006, Vol. 50, No. 10, pp. 778-784. 
Original Russian Text published in Astronomicheskii Zhurnal, 
2006, Vol. 83, No. 10, pp. 872-879.}

\maketitle

\begin{abstract}
The presence of ultra-high-energy cosmic rays (UHECR) results in an increase 
in the degree of ionization in the post-recombination Universe, which 
stimulates the efficiency of the production of H$_2$ molecules and the 
formation of the first stellar objects. As a result, the onset of the 
formation of the first stars is shifted to higher redshifts, and the 
masses of the first stellar systems decrease. As a consequence, a
sufficient increase in the ionizing radiation providing the reionization 
of the Universe can take place. We discuss possible observational 
manifestations of these effects and their dependence on the parameters of 
UHECR.
\end{abstract}


\section{INTRODUCTION}

\noindent

One of the scenarious \cite{berez,kuzmin,birkel} for the formation of 
ultra-high-energy cosmic rays (UHECR, with energies
above the Greisen--Zatsepin--Kuzmin cut-off, 
$E>10^{20}$~eV, \cite{greisen,zk}) suggests they are formed due to
the decay of ultra-heavy X particles with masses
$\geq 10^{12}$~GeV. In the interaction of UHECR with
low-energy cosmic-microwave background (CMB)
photons and the subsequent electromagnetic cascades,
resonance and ionizing photons are emitted,
which can increase fractional ionization of the
matter in the post-recombination Universe at redshifts
$\sim 10-50$ by factors of 5--10 compared to the
standard recombination regime. This circumstance
could qualitatively change the entire subsequent
evolution of the Universe, since the degree of ionization
substantially influences the rate of radiative
cooling by gas, and hence, the formation of the first
stellar objects. Indeed, cooling of the primordial gas at low
temperatures $T<10^3$~K is provided by thermal radiation
in rotational lines of H$_2$. In turn, H$_2$ molecules can
form in the primordial gas only via ion-molecule reactions
involving H$^-$ and H$_2^+$, in which electrons and
protons play the role of catalysts \cite{saslaw}.

Thus, the presence of UHECR in the early
Universe can appreciably influence the subsequent
stellar phase of its evolution. For this reason,
observational manifestations connected with characteristic 
features 
this phase can be used to constrain the parameters
of UHECR. In the present paper, we show that
due to the influence of UHECR stellar evolution in the Universe begins at earlier
epochs, and discuss
possible observational consequences of this prediction. 
Section 2 describes the ionization and
molecular kinetics of baryons in dark halos and their
thermodynamics in the presence of cosmic rays.
In Section 3, we present and discuss our results.
A summary is given in Section 4. In calculations we 
assumed a $\Lambda$-CDM model for the Universe:
$(\Omega_0,\Omega_{\Lambda},\Omega_m,\Omega_b,h ) = 
(1.0,\ 0.71,\ 0.29,\ 0.047,\ 0.72 )$
and deuterium abundance 
$n[D]/n = 2.6\times 10^{-5}$ \cite{spergel}.


\section{COSMIC RAYS AND THE THERMO-CHEMICAL EVOLUTION OF BARYONS IN DARK HALOS}

\noindent

Interaction of UHECR with CMB photons
produces high-energy particles -- photons, electrons,
positrons, and neutrinos, which transform through  
electromagnetic cascades into
ionizing Ly$c$ and resonant Ly$\alpha$ photons at the rate (per unit volume)
\cite{DN,peebles} 

\be
\label{eff}
{dn_{i,r} \over dt} = \epsilon_{i,r}(z)H(z)n. 
\ee
Here, $H(z)$ is the Hubble constant, $n$ is the baryon
density,

\begin{equation}
\label{param}
\epsilon_{i,r} \simeq {2.5 \times 10^{-4} \over 1+z} {M_{16}}^{2-\alpha} \Theta_{tot},
\end{equation}
is the efficiency of production of ionizing and resonant
photons \cite{DN}, where $\alpha$ is the spectral index of the
photon spectrum produced in the decay of 
a superheavy particle, $M_{16} = M_X/10^{16}~$GeV, $M_X$
is the mass of the superheavy particle, $\Theta_{tot}$ is a
function determined by the rate of decay of superheavy
particles at the present epoch and its redshift
dependence \cite{DN,sigl}. For the cosmic rays to produce measurable
distortions of the CMB radiation, the function
$\Theta_{tot}$ must be of the order of $\simeq 10^4{M_{16}}^{-0.5}$ \cite{DN}.
Following \cite{dnnn}, we assume that the efficiencies of
producing resonant and ionizing photons are similar:
$\epsilon_r(z) = \epsilon_i(z) =  \epsilon/(1+z)$.

In the standard model of the Universe, in the post-recombination
period ($z \simeq 1100$) before the formation
of the first stars ($z\simeq 30-20$), there are no sources
of ionizing photons. In these conditions, fractional ionization $x=n[\rm{H}^+]/n$
is determined by photo-recombinations, 
and ionization by additional Ly$c$ and Ly$\alpha$ photons
produced by the cosmic rays 

\be
\label{ionst}
\dot x = - k_1 n x^2 + {\epsilon \over 1+z} H(z)(1-x).
\ee
Under such conditions, the rate of formation of
molecular hydrogen, $f=n[\rm{H}_2]/n$, is given by the
expression

\begin{equation}
\label{h2}
\dot f = k_m n(1-x-2f)x,
\end{equation}
where

\be
\label{km}
k_m={k_2k_3\over k_3+k_4/(1-x)n}+{k_5k_6\over k_6+k_7/(1-x)n},
\ee
is the reduced rate of convergence of H into H$_2$ \cite{T97}
and $k_i$ are the rates of intermediate reactions
(see the table). At low temperatures ($T < 200$~K),
HD molecules can substantially influence the thermal
state of the gas \cite{varsh,shchekinov}. The kinetics of these
molecules is described by the equation

\begin{equation}
\label{hd}
\dot g = k_{\rm{D1}} f x n d_c - n x (k_{\rm{D1}} f + k_{\rm{D2}}) g,
\end{equation}
where $g=n[\rm{HD}]/n$ is the number density of HD
molecules, $d_c = n[\rm{D}]/n$ is the deuterium abundance,
and $k_{\rm{D1}}$ and $k_{\rm{D2}}$ are the rates of formation and destruction
of HD \cite{galli}. 

The growth of initial perturbations
in the dark matter leads to the formation of
gravitationally bound objects -- so called dark halos,
in whose potential wells baryons come into virial
equilibrium, and later on cool radiatively and give birth
to stars.

The role of UHECR at the initial star-formation
stage is determined by their influence on the thermal state
of the baryons. In order to understand this role, we consider
the evolution of baryons in a virialized halo 
using the simple system of equations (see, for instance, \cite{t93}) 

\begin{equation}
\label{radius}
\dot R= u,
\end{equation}

\begin{equation}
\label{velocity}
\dot u = {4kT\over \mu m_p R}-{4\over 3}\pi G(\rho_d+\rho)R,
\end{equation}

\begin{equation}
\label{temperature}
\dot T=-{2kT\over 3\mu R}u - \Sigma\Lambda_i,
\end{equation}
where $n=\rho/\mu m_p$, $\rho_d$ is the density of dark matter, $R$
is the radius of the region occupied by baryons, $u$
is the velocity of the boundary of the baryon cloud, 
$\Lambda_i$ are the cooling heating rates due to Compton
interactions with the CMB photons and radiation
in lines of atomic and molecular hydrogen and HD
molecules (expressions for the cooling and heating
rates are given in \cite{puy}). The initial values of the radius
and temperature were assumed to be equal to the
virial values for a given halo mass. The number
densities of electrons and of H$_2$ and HD molecules
were calculated for the virialization time, assuming a
simple prescription for the evolution of the density perturbation \cite{T97}.

\section{RESULTS}

Figure 1 shows the $z$ dependences of the temperature,
density, and Jeans mass of the gas for three
halos with the masses of $10^6\msun$, $10^7\msun$, and $10^8\msun$ 
that have reached a virial state at redshift $z = 20$, 
and for three values of the production efficiency of ionizing photons 
from cosmic rays ($\epsilon=0,~0.1,~1$). As an
example, let us consider in more detail the evolution
of a $10^7\msun$ halo. In the model with cosmic rays,
the number density of electrons at the virialization
time increases by approximately a factor of five. As a
result, the number density of H$_2$ molecules substantially
increases: during the contraction of the baryons,
it is $4\times 10^{-4}$ at $z = 16.5$ for the model with $\epsilon=0$,
while for $\epsilon=0.1$ it is already $3\times 10^{-3}$. As a result,
in the models with non-zero $\epsilon$, the gas temperature
falls faster, and becomes lower for lower densities.
A fairly high abundance of molecular hydrogen already
at the beginning of the contraction, together
with low temperature, favore 
binding almost all the deuterium in HD molecules,
due to the effects of chemical fractionation \cite{varsh}. When
$T<200$~K, the main input to the cooling process is
provided by HD molecules, and the cooling
is then so strong that the temperature drops
rapidly, and attains the CMB temperature, 
$\sim 50$~K. HD molecules provide
efficient heat exchange between the CMB radiation
and baryons via absorption of CMB photons and
subsequent transfer of the excitation energy to the
gas in collisional processes \cite{varsh}. As a result, the baryon
contraction becomes isothermal. The gas density 
grows rather rapidly, and starting from the 
value $\geq 3\times 10^7~\rm{cm^{-3}}$, the optical depth in the HD
lines exceeds unity. The Jeans mass varies as 
$M_J \sim 100 (1+z)^{3/2}n^{-1/2}\msun$, and reaches $\sim 1-2\msun$.
At higher densities, three-particle collisions become
the major mechanism for forming H$_2$, and at  
$n \sim 10^{8-9}$~cm$^{-3}$ the abundance of H$_2$ becomes
$\sim 1$. Thus, the central regions of the cloud become
opaque in H$_2$ lines \cite{palla83,palla86}.

Halos with lower masses evolve much 
slower, but as the efficiency of producing ionizing
photons increases, the abundances of electrons
and H$_2$ and HD molecules also increase, the temperature
decreases substantially, and the halo begins
to contract isothermally but at higher temperatures.
For example, in the model without cosmic rays, the
temperature of a $10^6\msun$ halo in this regime is 
$T\sim 300$~K, which can be explained by the less efficient
formation of H$_2$ and HD molecules. Thus, the transition
of a halo to isothermal contraction is
determined by its mass.

Let us consider the dependence of the redshift $z_t$
at which the transition to isothermal contraction
occurs, on the halo-virialization epoch $z_v$ and
the efficiency $\epsilon$, using as an example a halo with mass
$M_{3\sigma}$ and with a perturbation amplitude corresponding to
the $3\sigma$ level (see, for example, \cite{BL}). Figure 2
shows the dependence $z_t(z_v)$ for $\epsilon = 0,~0.1,~1$. The
upper $x$ axis shows the masses of perturbations with
amplitudes exceeding $3\sigma$ that result in the formation
of halos with mass of $M_h = M_{3\sigma}$ at $z_v$; the formation
of halos with $M_h > M_{3\sigma}$ has low probability. It is 
seen that for larger $\epsilon$ the halos evolve more
rapidly to the isothermal regime, particularly in the 
low-mass end. As the halo mass increases, 
the effect of decrease of the time needed to attain the
isothermal state weakens, since cooling and
contraction to high densities are possible in massive
halos due to energy losses in lines of atomic hydrogen. 

Thus, the presence of UHECR substantially accelerates
the halo evolution. With increasing $\epsilon$, the
contraction of the gas proceeds much more rapidly:
with $\epsilon = 0$, the baryons in a $10^7\msun$ halo 
virialized at $z = 20$ converge to isothermal
contraction at $z\approx 13.5$, while for $\epsilon = 3$ 
this transition occurs  
at $z\approx 17$. The influence of cosmic rays
on the thermal evolution can play a principal role for lower-mass
halos. For instance, when $\epsilon=0$, halos with 
$M_h \simeq 3\times 10^6\msun$ attain an isothermal state only by $z \sim 7$,
when the flux of external UV photons produced by the
stellar populations of more massive halos becomes
sufficiently strong for the temperature of the baryons in such
low-mass halos to exceed the virial temperature, and
for star formation to be suppressed \cite{BL,ferraraH2}. However,
when $\epsilon = 0.1$, such halos converge to the 
isothermal regime much earlier -- at $z \simeq 12$, when the
influence of the ionizing radiation from massive halos
is negligibly small. Large values of $\epsilon$ result not
only in a general acceleration of the evolution of halos
of all masses, but also in a shift of the minimum
mass $M_{\rm min}$ for the most rapidly evolving halos; this
mass corresponds to the maxima of the curves for different values of $\epsilon$
values in Fig. 2. When $\epsilon \simeq 1$, the gas in halos with 
masses $5\times 10^6\msun$ attains the isothermal
regime and conditions favorable for star formation already by
$z \approx 15$ (Fig. 2), i.e. long before the influence of earlier
formed low-mass halos can become important, while
larger-mass objects form later, so that their radiation
cannot affect the evolution of lower-mass
halos. Thus, the influence of UHECR is critical for
low-mass halos: in the presence of UHECR, the
masses of halos in which star formation is possible
decrease and a larger baryon fraction of 
the Universe becomes involved into stellar evolution.
Further, their stellar population can substantially change
the ionization and thermal state of the gas, 
since in the hierarchical scenario
of galaxy formation low-mass objects contain a larger fraction of the mass. 
This circumstance
may turn out to be of a fundamental importance for 
interpretation of the Wilkinson Microwave
Anisotropy Probe (WMAP) data \cite{spergel}.

According to the WMAP measurements of the
background polarization, the optical depth of the Universe
to Thomson scattering is $\tau_e = 0.16$ \cite{wmap}. Such
a high value corresponds to the beginning of the
reionization of the Universe at a redshift of
$z = 17$ \cite{wmap}. On the other hand, the spectra of distant
quasars show that reionization was completed only
by $z\sim 6$ \cite{becker,fan}. In addition, the flux of ionizing UV
radiation expected from quasars and young galaxies is
insufficient to reionize the Universe, even if the mass
distribution of stars, i.e., the initial mass function in the early Universe, was
skewed towards higher masses.
Additional sources of ionizing photons, such
as the radiation of accreting black holes \cite{cen,volon}, radiation
from microquasars \cite{cen03}, and the decay of unstable
neutrinos \cite{Sciama82,Sciama90,HH}, have been recently
considered to explain the early reionization. In these conditions,
the increase of the fraction of baryons condensed into 
stellar phase at the initial star formation
stage in the Universe, $f_*$, is an important
factor in providing the necessary rate of reionization.
In the presence of UHECR, this fraction certainly
grows, due to the increase in the efficiency of radiative
cooling of the baryons. Figure 3 shows the dependence of 

\be
f_* = {\int_{M_{\rm min}(z,\epsilon)}^{\infty}\psi(M)dM \over 
\int_{M_{\rm min}}^{\infty} \psi(M)dM}
\ee
on $\epsilon$, where $\psi(M)$ is the fraction of baryons present
in stars within mass $M$, calculated using the formalism
\cite{ps} [see (\ref{dndm}) below], $M_{\rm min}$ is the lower limit of 
masses of dark halos, and $M_{\rm min}(z,\epsilon)$ is the minimum mass of 
the dark halos that are able to cool and form
stars. The corresponding $z$ dependence of the number
of ionizing photons $f_{uvpp}$ per baryon associated with
stellar nucleosynthesis is shown in Fig. 3. It is seen from 
this dependence that at the supposed
epoch of reionization of the Universe ($z\leq 15$),
the number of ionizing photons in models with $\epsilon \neq 0$
can increase by half an order of magnitude. Thus, the full reionization of
the Universe may be determined by the stimulating
effect of UHECR on characteristics of the initial
stage of stellar nucleosynthesis.

The presence of UHECR in the early Universe
may be revealed from the properties of the characteristic
molecular-line emission arising in the stages
preceding formation of the first stars. At the prestellar
stage, dense and cold gaseous condensations
and molecular clouds form in dark halos, with 
substantial energy losses in H$_2$ and HD lines.
The total energy released in rotational lines of H$_2$
and HD may be $\epsilon_{\rm H_2} \sim 10^{12}$erg~g$^{-1}$. The total emission
from such molecular clouds in galaxies with masses
$10^9-10^{11}\msun$ can be substantial, and sufficient to
be detected by the planned infrared and submillimeter
telescopes ALMA\footnote{http://www.eso.org/projects/alma/}, ASTRO-F\footnote{http://www.ir.isas.ac.jp/ASTRO-F} and
SAFIR\footnote{http://safir.jpl.nasa.gov}. The
influence of UHECR will be manifested as an increase
in the relative specific energy release in H$_2$ and HD lines 

\be
\label{en}
<E_{\rm H_2}> = {\int_{M_{min}(z,\epsilon)}^{\infty}\epsilon_{\rm H_2} \psi(M)dM \over \int_{M_{min}}^{\infty}\psi(M)dM },
\ee
as well as an increase in the redshift at which this
emission is detected, $z_{\rm H_2}$. This dependence is shown
in Fig. 4. An unambiguous relationship between the
observable quantities $\langle E_{\rm H_2}\rangle$ and $z_{\rm H_2}$ and the parameter
$\epsilon$ is evident.

Another physical parameter of the initial stage of
stellar evolution that will become accessible to observations
in the nearest future is the supernova rate.
One of the key projects of the James Webb Space
Telescope\footnote{http://www.stsci.edu/jwst/} 
is to observe supernovae at high redshifts,
$z \sim 10-20$. Since at least one supernova may explode
in every low-mass halo \cite{AbelN02}, the number of supernovae
at high redshifts should increase if the minimum
halo mass is decreased due to the influence of
cosmic rays. Figure 5 shows the redshift dependence
of the expected specific (per unit mass) supernova
rate in the Universe for various values of $\epsilon$, calculated
for a standard hierarchical model of galaxy formation  within 
the formalism \cite{ps}, for the spectrum of
perturbations with $n = 1$. The differential mass distribution
of the dark halos at a given redshift is

\be
\label{dndm}
{dn \over dM}(M,z)dM = \sqrt{2\over \pi}{\delta_c \over D(z) \sigma^2(M)}{d\sigma(M) \over dM}\exp{-{\delta_c}^2\over 2D^2(z)\sigma^2(M)}dM
\ee
where $D(z)$ is the perturbation growth factor, $\delta_{c}=1.69$
is the density parameter, and $\sigma(M)$ is the rms
deviation of the perturbations inside mass $M$. Summing over
mass in (\ref{dndm}) gives the number of objects
per unit comoving volume at redshift $z$. Estimates
of the number of supernovae in a halo are based on
currently existing models \cite{snrate} 

\be
\label{snrates}
\gamma(z) = {\nu \Omega_b f_b \over \tau t_{ff}}
 \simeq 1.2 \times 10^{-7}\Omega_{b,5} f_{b,8} (1+z)^{3/2}_{30} M_6~yr^{-1},
\ee
where ${(1+z)}_{30} = (1+z)/30$ and $M_6= M/10^6\msun$.
A Salpeter initial mass function is assumed, which
corresponds to one supernova per every $56\msun = \nu^{-1}$, 
$\Omega_b=0.05\Omega_{b,5}$, and $f_b = 0.08 f_{b,8}$ is the 
fraction of baryons which can cool and form stars \cite{abel98}. 
The density of dark matter is
$\rho = 200\rho_c = 200[1.88\times10^{-29}h^2(1+z)^3]$, and the
free-fall time is $t_{ff}=(4\pi G\rho)^{-1/2}$. The efficiency of
star formation was normalized to the Galactic value,
${\tau}^{-1} = 0.6\%$. It is readily seen that at high redshifts, 
($z\geq 15$), the number of supernovae grows approximately
linearly with $\epsilon$.

\section{CONCLUSIONS}
\noindent

We conclude that the increase of the fractional  
ionization in the post-recombination Universe
caused by ultra-high energy cosmic rays, 
has a stimulating effect on the initial stages of stellar
evolution in the Universe: the epoch of the formation
of the first stars is shifted to higher redshifts, and the
minimum mass for systems in which star formation is
possible decreases. This should be manifested in several
different observational effects sensitive
to the efficiency of UV photon production by
the cosmic rays, $\epsilon$. Considered together, these effects
should make it possible to infer the parameters
of the UHECR theory or obtain additional constraints
for them. These manifestations include emission 
in rotational lines of H$_2$ and HD irradiated by contracting protostellar
condensations, and 
the supernova rate and its dependence on redshift.
Morevover, enhancement of the initial stage of star
formation by cosmic rays can produce additional
ionizing photons, which seem currently to be lacking for 
providing efficient reionization of the Universe.



\newpage

\begin{table}[!h]
\caption{
Chemical-reaction rates.
}
\center
\bigskip
\begin{tabular}{|l|l|c|}
\hline
Reaction      &Reaction coefficient  $k$ $\rm [cm^3 s^{-1}]$     &Reference\\
\hline
${\rm H^+ +e^- \to H +h\nu}$  &$k_1\approx 1.88\tento{-10} T^{-0.64}$  &\cite{hutchins}\\
${\rm H+e^- \to H^-+h\nu}$   &$k_2\approx 1.83\tento{-18} T^{0.88}$     &\cite{hutchins}\\
${\rm H^-+H \to H_2 + e^- }$       &$k_3\approx 1.3\tento{-9}$          &\cite{hirasawa}\\
${\rm H^++H \to H_2^+ +h\nu }$  &$k_5\approx 1.85\tento{-23}T^{1.8}$   &\cite{shapiro}\\
${\rm H_2^++H \to H_2+H^+ }$       &$k_6\approx 6.4\tento{-10}$       &\cite{karpas}\\
${\rm D^+ +H_2 \to HD +H^+ }$   &$\alpha_1\approx 2.1 \times 10^{-9} $   &\cite{galli}\\
${\rm H^+ +HD \to H_2 +D^+ }$ &$\alpha_2\approx \alpha_1 e^{-465 \ K/T}/4$ &\cite{galli}\\
$H^-+h\nu \to H + e^-$      &$k_4\approx 0.114\Tg^{2.13}e^{-8650/\Tg}$  &\cite{galli}\\
$H_2^+ +h \nu \to H + H^+$   &$k_7\approx {6.36}{5} e^{-71600/\Tg}$     &\cite{galli}\\
\hline
\end{tabular}%
\label{RateTable}
\end{table}

\newpage

\twocolumn


\begin{figure}
\epsfxsize=8cm
\epsfbox{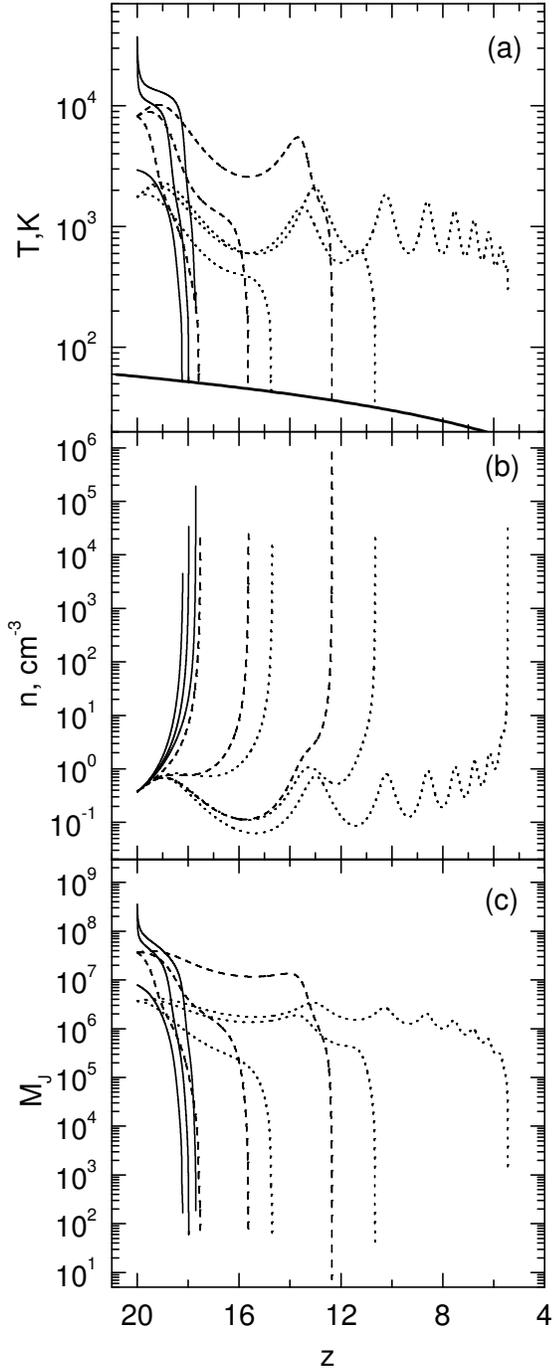}
\caption{
Evolution of the (a) temperature, (b) density, and
(c) Jeans mass for three halos with masses of $10^6$
(dotted), $10^7$ (dashed), and $10^8$ (solid) that have
reached a virial state at redshift $z = 20$, and for three
efficiencies of the production of ionizing photons from
cosmic rays, $\epsilon = 0,~0.1,~3$  (from
left to right). The solid curve in the upper plot
shows the temperature of the CMB.
}
\label{evol}
\end{figure}

\begin{figure}
\epsfxsize=8cm
\epsfbox{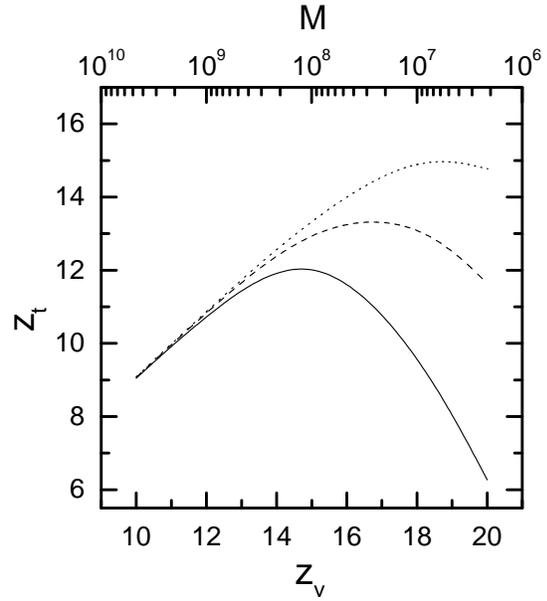}
\caption{
Dependence of the redshift $z_t$ for the transition to
a state of isothermal gas contraction on the epoch of halo
formation $z_v$ and the efficiency $\epsilon$ (the solid, dotted, and
dashed curves correspond to $\epsilon = 0$, $\epsilon = 0.1$, and $\epsilon = 1$,
respectively). The plot shows the evolution of a halo with
mass $M_{3\sigma}$ corresponding to $3\sigma$ perturbation emerged at $z_v$.
}
\label{accel}
\end{figure}

\begin{figure}
\epsfxsize=8cm
\epsfbox{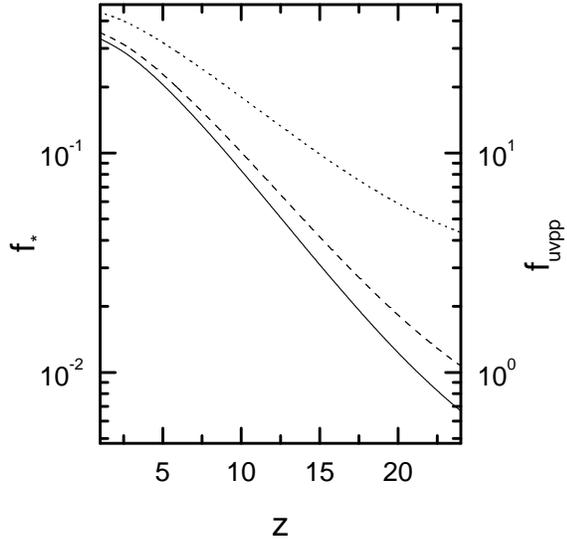}
\caption{
Redshift dependence of the fraction of baryons
condensed to a stellar phase in the
early star-formation stage in the Universe, $f_*$, for $\epsilon = 0,~0.1,~1$ (solid, dashed, and dotted, respectively). A standard
model for hierarchical clustering is assumed. The
scale in the right shows the mean number of ionizing 
(UV) photons emitted by the stellar population per
baryon, $f_{uupp}$.
}
\label{fstar}
\end{figure}

\begin{figure}
\epsfxsize=8cm
\epsfbox{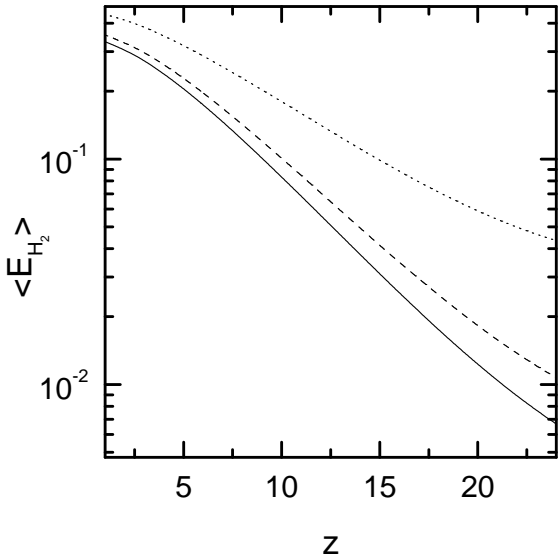}
\caption{
Redshift dependence of the relative specific energy
released in H$_2$ and HD lines for $\epsilon = 0,~0.1,~1$ (solid, dashed,
and dotted, respectively). A standard model for hierarchical
clustering is assumed.
}
\label{aveemis}
\end{figure}

\begin{figure}
\epsfxsize=8cm
\epsfbox{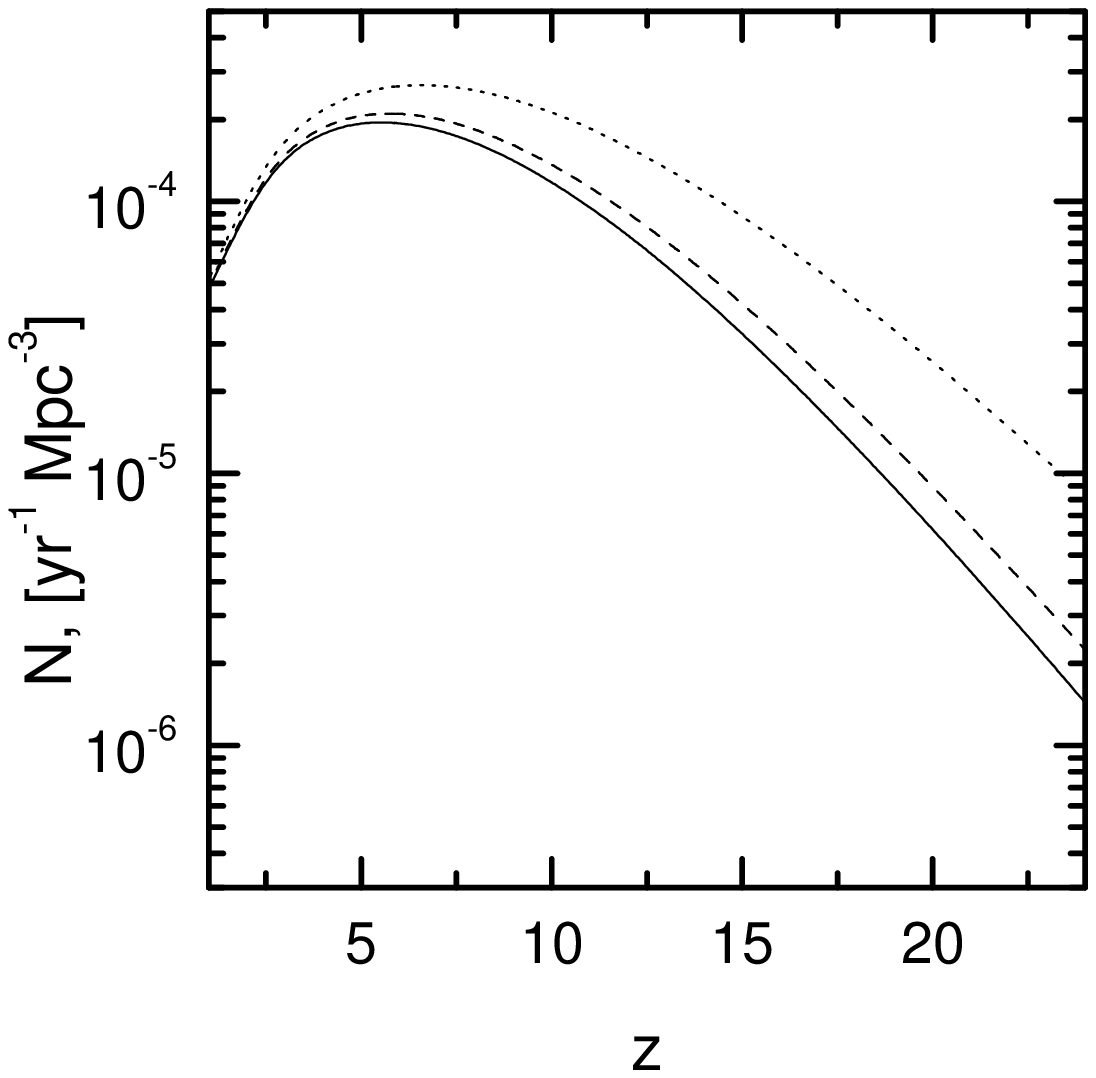}
\caption{
Redshift dependence of the supernova rate for
$\epsilon = 0,~0.1,~1$ (solid, dashed, and dotted, respectively). A
standard model for hierarchical clustering is assumed.
}
\label{snr}
\end{figure}


\begin{thebibliography}{99}

\bibitem{berez} V.S. Berezinsky, M. Kachelrie\ss \ and A. Vilenkin, Phys. Rev. Lett. {\bf 79}, 4302 (1997).

\bibitem{kuzmin} V.A. Kuzmin and V.A. Rubakov, Phys. Atom. Nucl. {\bf 61}, 1028 (1998).

\bibitem{birkel} M. Birkel and S. Sarkar, Astropart. Phys. {\bf 9}, 298 (1998).

\bibitem{greisen} K. Greisen, Phys. Rev. Lett. {\bf 16}, 748 (1966).

\bibitem{zk} G.T. Zatsepin, V.A. Kuzmin, Pis'ma v JETPh {\bf 4}, 114 (1966), [JETP Lett. 4, 78 (1966)].

\bibitem{DN} A.G. Doroshkevich and P.D. Naselsky, Phys. Rev. D, {\bf 12}, 123517 (2002).

\bibitem{dnnn} A.G. Doroshkevich, I.P. Naselsky, P.D. Naselsky, I.D. Novikov, Astrophys. J. {\bf 586}, 709 (2003).

\bibitem{saslaw} W.C. Saslaw, and D. Zipoy, Nature {\bf 216}, 967 (1967).

\bibitem{spergel} D. N. Spergel, L. Verde, H.V. Peiris et al., Astophys. J. Suppl. {\bf 148}, 175 (2003).

\bibitem{peebles} P.J.E. Peebles, S. Seager, and W. Hu, Astrophys. J. {\bf 539}, L1 (2000).

\bibitem{sigl} P. Bhattacharjee, and G. Sigl, Phys. Rept. {\bf 327}, 109 (2000).

\bibitem{T97} M. Tegmark, J. Silk, M.J. Rees, et al.,
Astophys. J. {\bf 474}, 1 (1997).

\bibitem{hutchins} J.B. Hutchins, Astrophys. J. {\bf 205}, 103 (1976). 

\bibitem{hirasawa} T. Hirasawa, Progr. Theor. Phys. {\bf 42}, 523 (1969).

\bibitem{shapiro}  P.R. Shapiro and H. Kang, Astrophys. J. {\bf 318}, 32
(1987). 

\bibitem{karpas} Z. Karpas, V. Anicich and W.T. Huntress, J. Chem. Phys.
{\bf 70}, 2877 (1979).

\bibitem{galli} D. Galli and F. Palla, Astron. and Astropys. {\bf 335}, 403 (1998).

\bibitem{varsh} D. A. Varshalovich and V. K. Khersonskii, Pis’ma
Astron. Zh. 2, 574 (1976) [Sov. Astron. Lett. 2, 227 (1976)].

\bibitem{shchekinov} Yu. A. Shchekinov, Pis’ma Astron. Zh. 12, 499 (1986)
[Sov. Astron. Lett. 12, 211 (1986)].

\bibitem{t93} M. Tegmark, J. Silk, A. Evrard, Astrophys. J. {\bf 417}, 54 (1993).

\bibitem{puy} D. Puy and M. Signore, NewA {\bf 3}, 247 (1998).

\bibitem{palla83} F. Palla, S.W. Stahler and E.E. Salpeter, Astrophys. J. {\bf 271}, 632 (1983).

\bibitem{palla86} S.W. Stahler, F. Palla and E.E. Salpeter, Astrophys. J. {\bf 302}, 590 (1986).

\bibitem{BL} R. Barkana, and A. Loeb, Phys. Rept. {\bf 349}, 125 (2001).

\bibitem{ferraraH2} A. Ferrara, Astrophys. J. {\bf 499}, L17 (1998).

\bibitem{wmap} A. Kogut, D.N. Spergel, C. Barnes, et al., Astrophys. J. Suppl. 
{\bf 148}, 161 (2003).

\bibitem{becker} R. Becker, X. Fan, R.L. White et al., Astron. J. {\bf 122}, 2850 (2001).

\bibitem{fan} X. Fan, V.K. Narayanan, M.A. Strauss et al., Astron. J. {\bf 123}, 1247 (2002). 

\bibitem{cen} R. Cen, Astrophys. J. {\bf 591}, 12 (2003).

\bibitem{volon} P. Madau, M.J. Rees, M. Volonteri et al., 
Astophys. J. {\bf 604}, 484 (2004)

\bibitem{cen03} R. Cen, Astrophys. J. {\bf 591}, L5 (2003).

\bibitem{Sciama82} D.W. Sciama, Mon. Not. R. Astron. Soc. {\bf 198}, 1 (1982).

\bibitem{Sciama90} D.W. Sciama, Phys. Rev. Lett. {\bf 65}, 2839 (1990).

\bibitem{HH} S.H. Hansen and Z. Haiman, Astrophys. J. {\bf 600}, 26 (2004).

\bibitem{ps} W.H. Press and P. Schechter, Astrophys. J., {\bf 187}, 425 (1974).

\bibitem{AbelN02} T. Abel, G.L. Bryan and M. Norman, Science {\bf 295}, 93 (2002).

\bibitem{snrate} S. Marri, A. Ferrara, L. Pozzetti, Mon. Not. R. Astron. Soc. 
{\bf 317}, 265 (2000).

\bibitem{abel98} T. Abel, P. Anninos, Y. Zhang and M.L. Norman, Astrophys. J.,	
{\bf 508}, 518 (1998)


\end{thebibliography}
\end{document}